%% file: quinngpu11-D.tex
\documentclass[preprint]{elsarticle}
\usepackage{graphicx}
\usepackage{amsmath}
\usepackage{bm}

\include{definitions}

\begin{document}
\title{Data Assimilation using a GPU Accelerated Path Integral Monte Carlo Approach}
\author{John C. Quinn}
\address{ Department of Physics\\
and\\
BioCircuits Institute\\}
\ead{jquinn@ucsd.edu}
\author{Henry D.I. Abarbanel}
\address{Department of Physics,\\
Marine Physical Laboratory (Scripps Institution of Oceanography)\\
and\\
Center for Theoretical Biological Physics\\}
\ead{habarbanel@ucsd.edu}
\address{University of California, San Diego\\
La Jolla, CA 92093-0402  USA\\}

\begin{abstract}
The answers to data assimilation questions can be expressed as path integrals over all possible state and parameter histories.  We show how these path integrals can be evaluated numerically using a Markov Chain Monte Carlo method designed to run in parallel on a Graphics Processing Unit (GPU).  We demonstrate the application of the method to an example with a transmembrane voltage time series of a simulated neuron as an input, and using a Hodgkin-Huxley neuron model.  By taking advantage of GPU computing, we gain a parallel speedup factor of up to about 300, compared to an equivalent serial computation on a CPU, with performance increasing as the length of the observation time used for data assimilation increases. 
\end{abstract}


\maketitle

\section{Introduction}

Data assimilation is designed to utilize information from measurements of an observed physical or biological system to complete a model of that system. Completion means accurately estimating unknown parameters in the model and accurately estimating the unobserved model state variables at the end, time $T$, of the observation window. The information on the parameters $\p$ and the full set of model state variables $\x_T$ allow forecasting the state of the system using the model. Also knowledge of the state $\x_T$ permits assessing the quality of the model.

Typically the model will take the form of a system of nonlinear differential equations, either based on the underlying physics of the system in question, or designed to reproduce observed phenomena. To make quantitative predictions the parameters of the model need to be adjusted to fit the data.  Since neither the model nor the measurements are exact, the problem should be treated probabilistically.  The knowledge we gain from the observations allows us to estimate the conditional probability distribution $P(\x_T|\y_{1:M})$, where $\y_{1:M}=\{\y_1,\y_2,\ldots,\y_M\}$ is the collection of observations up to time $t_M = T$. Moments and marginal distributions of parameters or state variables may also be estimated.

Within an observation window, one makes measurements at time $t_n = \{t_1,t_2,\ldots,t_M = T\}$. At each time $t_n$, $L$ observations are made, represented by the $L$-dimensional vector $\y(t_n) = \y_n$. The observations provide information about the system state, but it is not complete information, because the observations are likely to be noisy, and more importantly, the dimension $L$ of $\y$ is typically less than the dimension of the system state $x_{n,a};\;a=1,2,\ldots,D \ge L$.  

The problem of data assimilation was formulated in a probabilistic way some time ago~\cite{cox64, friedland66}.  The  key question is this: given a time series of observations $\Y=\y_{1:M}=\{\y_1,\y_2,...,\y_M\}$ and a model, what model state histories $\X = \{\x_0, \x_1, \ldots, \x_M \}$ and set of parameters $\p =\{p_1,p_2,...,p_{N_p}\}$ could have produced the observed data? The conditional distribution over state histories and conditioned on observations is called $P(\X|\Y)$.

Various methods have been developed to solve this problem.  Kalman filtering~\cite{kalman60} is the exact solution when the model is linear and the noise is Gaussian and white in time.  A more general and computationally intensive approach is particle filtering~\cite{vanLeeuwen09, arulampalam, doucet}, where the distribution over states is approximated by an ensemble of particles.   A similar approach is to phrase the problem in terms of continuous time path integrals~\cite{eyink05, restrepo, apte07}. The continuous time path integrals are approximations, and in discrete time an exact formulation has been given~\cite{abar09}.  In the path integral approach, the whole time history of states $\X$ is considered at once, instead of sequentially calculating the states at discrete  time steps, $\x_n$.  

What we really want are moments of functions on the path $\X$ or histograms of the marginal distributions, $P(x_{n,l}|\Y)$,  where $x_{n,l}$ is one component of $\X$.  These can be expressed as path integrals over all possible state histories $\X$, with each path weighted by $P(\X|\Y)$.  The path integrals can then be approximated using standard Markov Chain Monte Carlo techniques.  The main difficulty with this approach is that the methods typically require very many iterations to get good results, and so they can be very computationally expensive.

This problem can be ameliorated by taking advantage of parallel computing technology using a graphics processing unit (GPU).  These computations can be done on inexpensive desktop computers, with an off-the-shelf GPU, which can execute hundreds of threads in parallel (see \cite{preis09} for another use of GPU computing in physics).  The Markov chain still must be updated sequentially of course, but the computations done on each iteration can be highly parallelized. Since in the path integral approach, all time steps are considered at once, it is possible to parallelize the algorithm such that each thread does the computations associated with a different time step. This leads to a large decrease in the amount of computation time required, and makes many problems become practical.  We show an example problem using simulated neuron voltage time series data, and the Hodgkin-Huxley neuronal model as inputs into the procedure.  In this example we get a parallel speedup factor of up to about 300.

\section{Path Integral Formulation of Data Assimilation}

This section is review of previous work \cite{cox64, friedland66, abar09, pham, jcq10} and an introduction of notation. The idea is to express the answer to probabilistic data assimilation questions as path integrals over all possible state histories, and all possible parameter values. 
  
The inputs into this formulation will be a model of the observed system and a time series of observations, together with uncertainties associated with these two ingredients.  The model will be in the form of Markov transition probabilities, which describe how the model state evolves in time.  The transition probabilities may come from a discrete time map or a system of ordinary differential equations which is then discretized.  In this section we will show how these ingredients are combined to form a probability distribution that is a function of the time history of the model state conditioned on the observations.  

The first ingredient is the model.  To make a model, we first represent the state of the system at each time step $n \in \{0,1,\ldots, M\}$ as a $D$-dimensional vector $\x_n$.  We assume the dynamical model is Markov, and so we can represent the time evolution using a transition probability $P(\x_{n+1}|\x_n)$ which depends only on $\x_{n+1}$ and $\x_n$, and not any previous states.  This tells us the probability of transitioning from state $\x_n$ at time step $n$ to state $\x_{n+1}$ at time step $n+1$.  If there is no noise in the model, this is a deterministic transition in which case $P(\x_{n+1}|\x_n)$ will be a delta function. When noise is present, typically due to a noisy environment or arising from resolution errors associated with discretization of the model in time and space, this will be represented as stochastic transition in which case $P(\x_{n+1}|\x_n)$ will a broader version of a delta function, with a width which depends on the noise level.  The $D$ dimensions of the model may include any time-independent parameters, by promoting the parameters to state variables with time evolution given by $\dot{\p} = 0$. 

The second ingredient is a time series of measurements, $\Y =\y_{1:M} = \{\y_1, \y_2, \ldots, \y_M\}$.  At each time step labeled by $n \in \{1,2,\ldots, M\}$ the measurement is an $L$-dimensional vector $\y_n$. Part of the model development consists of associating a known `observation' function $\h(\x_n)$ of the model state variables $\x_n$ with the observation $\y_n$. Usually one assumes that the observations and the observation function are related by additive noise added to the observation, so $\y_n \approx \h(\x_n) + \mbox{noise}$, but we first discuss a more general case. Typically $L<D$, so there are `hidden states' which must be inferred using all the information available from observations and from the model.  More keeping with a physicist's view of measurements, we refer to the $D-L$ unmeasured states as unobserved state variables. We need to estimate them as well as the fixed parameters. We will assume later that the observation only depends on the state at the current time step.

We then consider the discrete time evolution of the model state by defining a path variable $\X = \x_{0:M} = \{\x_0,\x_1, \ldots, \x_M\}$.  This describes a trajectory through the the $D$-dimensional state space.  This formulation considers the entire path simultaneously, assuming all the observations have already been collected.  In signal processing language this method would be called a `smoother', as opposed to a `filter' which only uses observations from the past.

We then use the Markov transition probabilities, $P(\x_n|\x_{n-1})$ specified by the model to find $P(\x_n|\y_{1:n-1})$ from $P(\x_{n-1}|\y_{1:n-1})$.  We also use Bayes' rule~\cite{mackay} to introduce the observations into the formulation.  The result is

\bea \label{RBE}
&&P(\x_n|\y_{1:n}) =  \frac{P(\y_n|\x_n, \y_{1:n-1})}{P(\y_n|\y_{1:n-1})} \int d\x_{n-1} P(\x_n | \x_{n-1}) P( \x_{n-1} | \y_{1:n-1}), \nonumber \\
&=&  \frac{P(\y_n,\x_n| \y_{1:n-1})} {P(\x_n|\y_{1:n-1}) P(\y_n|\y_{1:n-1})} 
\int d\x_{n-1} P(\x_n | \x_{n-1}) P( \x_{n-1} | \y_{1:n-1}) \nonumber \\
&=& e^{\mathrm{CMI}(\x_n,\y_n|\y_{1:n-1})}\int d\x_{n-1} P(\x_n | \x_{n-1}) P( \x_{n-1} | \y_{1:n-1}),
\eea
where we identify the factor in front of the integral as the exponential of the conditional mutual information
between the state $\x_n$ at time $t_n$ and the measurement $\y_n$ at that time. 

This is the key equation that is used by all recursive Bayesian estimation methods.  It is the starting point for particle filter methods \cite{arulampalam} which use this equation to perform the forecast step to advance the probability distribution, as approximated by a weighted particle distribution, and the analysis step to incorporate measurements.  Particle filtering methods process the measurements and update the model state sequentially.  Here we take a different approach and consider the entire model evolution and the whole time series of observations simultaneously.      

To write $P(\x_M|\y_{1:M})$ over a path in state space we first look at the initial  time step by setting $n=1$ into Eq.~(\ref{RBE}) to arrive at
\[
P(\x_1|\y_{1:1}) =  \frac{P(\y_1|\x_1)}{P(\y_1)} \int d\x_0 P(\x_1 | \x_0) P( \x_0),
\]
and we note that $\y_{1:0}$ means there are no measurements to condition the probabilities by.  We then apply Eq.~(\ref{RBE}) iteratively $M-1$ additional times to establish
\begin{equation}\label{PXM}
P(\x_M|\y_{1:M}) = \int \prod_{n=1}^{M} d\x_{n-1} \frac{P(\y_n|\x_n, \y_{1:n-1})}{P(\y_n|\y_{1:n-1})} P(\x_n|\x_{n-1}) P(\x_0).
\end{equation}
This is the marginal distribution of the state at the last time point $\x_M$.  

This is also the integral representation of the solution to an underlying Fokker-Planck equation
for the problem of a state $\x_n$ satisfying noisy dynamical equations being informed by noisy measurements $\y_n$.  When the noise is Gaussian the underlying Fokker-Planck equation is known as the Zakai equation \cite{zakai}.  

We can use the definition of a marginal distribution
\[
P(\x_M|\y_{1:M}) = \int d\x_0 \ldots d\x_{M-1} P(\x_{0:M}|\y_{1:M}),
\] 
to express the conditional probability distribution as a function of the entire path $\x_{0:M}$ via
\be \label{FullPathCPD}
P(\x_{0:M}|\y_{1:M}) =  P(\X|\Y) = \prod_{n=1}^{M}  \left[\frac{P(\y_n|\x_n, \y_{1:n-1})}{P(\y_n|\y_{1:n-1})} \right] P(\x_n|\x_{n-1}) P(\x_0).
\ee
To calculate the marginal distribution at a particular time step $0 \le n \le M$, $P(\x_n| \y_{1:M})$, we integrate over the state variables at all the other time steps.  Note that this formulation for $P(\x_n|\y_{1:M})$ includes the information gained from the whole time series of measurements $\y_{1:M}$ and not just the the past measurements $\y_{1:n}$.

We define the action, $A$, by taking the log of Eq.~(\ref{FullPathCPD}) and dropping additive terms independent of $\X$,
\be \label{action}
-A(\X|\Y) = \sum_{n=1}^M \log P(\y_n|\x_n, \y_{1:n-1}) + \sum_{n=1}^M \log P(\x_n| \x_{n-1}) + \log P(\x_0).
\ee
The first term is where the information from the measurements is incorporated, the second term is where the dynamical model is incorporated, and the last term is the prior distribution of initial states and parameters, which takes into account any previous assumptions.  Using this definition, we can write
\[
P(\X|\Y) \propto e^{-A(\X|\Y)}.
\]  
The constant of proportionality is independent of $\X$, so the expectation of any function $G(\X)$ on the path $\X$, conditioned on the measurements $\Y$, is written as
\be \label{pathInt}
E[G(\X)|\Y] =\, \left<G(\X) \right>\,=\frac{\int d\X\,G(\X)\,e^{-A(\X|\Y)}}{\int d\X\,e^{-A(\X|\Y)}}.
\ee 

\subsection{Assumptions leading to simplified form of action}
We assume that the measurement at time step $n$ depends in a known and deterministic way given by $\h$ on the current state, $\x_n$, plus multivariate Gaussian noise. We also assume that $\y_n$ is independent of any earlier measurement, and so
\begin{align}\label{obsError}
\y_n - \h(\x_n) &= \N(0, \Ro^{-1}) \notag\\
               &\equiv \de_n, 
\end{align}
where $\N(0,\Ro^{-1})$ represents a random vector drawn from a multivariate Gaussian with zero mean and covariance matrix $\Ro^{-1}$.  We these assumptions, the term inside the first sum of Eq.~(\ref{action}) becomes 
\[
P(\y_n| \x_n, \y_{1:n-1}) = P(\y_n|\x_n) \propto \exp\left[ -\frac{1}{2} \de_n^T \cdot \Ro \cdot \de_n \right].
\]

Next we make some assumptions about the dynamics. 
We define the model error term by by discretizing the model differential equation, $\frac{d\x}{dt} = \F(\x(t))$, using the trapezoid rule with time step $\Delta t$:
\begin{equation}\label{modelError}
\ep_n \equiv \x_{n} - \x_{n-1} - \frac{\Delta t}{2} \left[ \F(\x_{n}) + \F(\x_{n-1}) \right].
\end{equation}
If the dynamics were deterministic, then $\ep_n =0$  (exact in the limit where $\Delta t \to 0$) for all $n$.  We allow noise into the dynamics by relaxing this condition by replacing the zero with a stochastic term: 
\[
 \ep_n = \N(0, \Rd^{-1}). \notag\\
\]
We have allowed for some types of random error in the model by introducing a noise term, which is an additive Gaussian noise applied at every time step.  The noise is parameterized by the covariance matrix $\Rd^{-1} = \Delta t \mathbf{\Gamma}$, where $\mathbf{\Gamma}$ is a diffusion matrix.  We these assumptions, the term inside the second sum of Eq.~(\ref{action}) becomes
\[
P(\x_n| \x_{n-1}) \propto \exp\left[ -\frac{1}{2} \ep_n^T \cdot \Rd \cdot \ep_n \right].
\]

Finally, for simplicity we assume that $P(\x_0)$ is uniform inside the boundaries and zero outside.  This means the initial states and parameters can be restricted to fall inside a boundary, and that the last term in Eq.~(\ref{action}) can be dropped because it is a constant inside the boundary.  

The action can be written as $A = A_o + A_d$, where
\begin{align}
A_o &= \frac{1}{2}\sum_{n=1}^{M}  \de_n^T \cdot \Ro \cdot \de_n,  \notag \\
A_d &= \frac{1}{2} \sum_{n=1}^{M} \ep_n^T \cdot \Rd \cdot \ep_n, \notag
\end{align}
are the contributions from the observations and the dynamics respectively.

\section{Monte Carlo Evaluation}

Now that we have a formulation of $P(\X|\Y)$, we would like to calculate quantities such as means and covariances, which can be used to make estimates and predictions.  These quantities can be written as path integrals of the form given in Eq.~(\ref{pathInt}), where $G(\X)$ is chosen to be some function of the path that is of interest.  

The challenge is to evaluate these path integrals.  One way to do this is to  generate a series of paths $\{\X^{(1)},\ldots, \X^{(J)}\}$ that are distributed in path space according to $P(\X|\Y) \propto\exp[-A(\X|\Y)]$, then we can use those paths to approximate the distribution with
\[
P(\X|\Y) \approx \frac{1}{J} \sum_{j=1}^J \delta( \X - \X^{(j)}).
\]
We can then calculate expectation values of any function of the path with
\begin{align}\label{MCapprox}
 \left< G(\X) \right> &= \int d\X \;G(\X) P(\X|\Y) \notag \\
   &\approx \frac{1}{J} \sum_{j=1}^{J} G(\X^{(j)}).
\end{align}

The sample paths can be thought of as representing many possible time evolutions of the system state that could have produced the observed data when measured.  The paths that are more likely to have produced the observed data will be generated more times than the paths which are less likely to have produced the observed data.  

Now we need a method which will produce a series of paths that are distributed according to $\exp[-A(\X)]$.  There are several path integral Monte Carlo methods, such as Metropolis or Hybrid Monte Carlo, that do exactly this \cite{MRT, mackay, neal}.

\subsection{Metropolis Monte Carlo}
One of the simplest and oldest approaches is the Metropolis Monte Carlo method \cite{MRT}.  This method works by generating a sequence of paths  $\{\X^{(1)},\ldots, \X^{(J)}\}$ that we will call Monte Carlo paths by a random walk through path space.  The method is an example of a Markov Chain Monte Carlo method, because the paths are generated in sequence and the next path is generated only from the current path in a stochastic way. The random walk is biased in a particular way as described below so that the sequence of paths that are generated come out distributed according to $\exp[-A(\X)]$.

The method works by generating a new path $\X^{(n+1)}$ from the current path $\X^{(n)}$ with a two step procedure.  First, a candidate path  $\X'$ is proposed by adding an unbiased random displacement to the current path $\X =\X^{(n)}$.  The displacement may be to only one component or all the components, and may be drawn from any type of distribution, as long as it is unbiased; this assures that $\X \to \X'$ is as likely to occur as $\X' \to \X$. 

The proposal distribution actually does not need to be unbiased if the acceptance probability is modified in the way shown in \cite{hastings}.  There are several methods which make a smarter choice of the proposal distribution, such as the force bias method \cite{pangali}.  Such methods may be converge with fewer iterations, but they have the added complexity of requiring derivatives of the action to be computed.   

Next the proposed path is either accepted ($\X^{(n+1)} = \X'$) or rejected ($\X^{(n+1)} = \X$).  The probability for acceptance is 
\be \label{Paccept}
P_{accept}(\X', \X) = \min( 1, \exp[-\Delta A(\X',\X) ] ),
\ee
where $\Delta A(\X',\X) = A(\X') - A(\X)$ is the change in action which would be caused by changing $\X$ to $\X'$.  This says that if a proposed change will lower the action it should be accepted, and if the proposed change increases the action it should only be accepted with probability $\exp[-\Delta A(\X',\X)]$. Note that only the change in action is required, so the full action never needs to be computed.  This means we do not need to keep track of additive constants to the action, and can save a lot of computation time by only computing the terms in the action that will be changed by the update.

\section{Parallel Implementation for GPUs}

The Metropolis Monte Carlo method is simple and powerful, but it has the drawback of requiring very many path updates to get accurate statistics.  Since so many path updates are required, the computation becomes very expensive in terms of computer time.  One way to deal with this problem is to take advantage of parallel computing technology, using a Graphics Processing Unit (GPU).  With GPU technology, and using Compute Unified Device Architecture (CUDA), it is possible to execute hundreds of threads of execution simultaneously.  Typically each thread will perform the same operations, but on different pieces of the data.  Of course since the paths are updated sequentially, the path update process cannot be run in parallel.  However, the many computations needed on each iteration can be done in parallel by having different threads work on different time steps.

\begin{figure}
\begin{enumerate}
\item For each odd $n$, a thread does the following:
\begin{enumerate}
 \item Proposed change: $\x_n' = \x_n + \bf{\Delta}_n \cdot \mathbf{U}(-1,1)$
 \item Calculate and store new RHS of model equations: $\F_n' = \F(\x_n')$
 \item Calculate model error terms: $\ep_n, \ep'_n, \ep_{n+1}, \ep_{n+1}'$
 \item Calculate observation error terms: $\de_n, \de_n'$
 \item Change in dynamical part of action: \\
 $\Delta A_{d,n} = \frac{R_d}{2} \left[ (\ep_n')^2 - \ep_n^2 + (\ep_{n+1}')^2 - \ep_{n+1}^2 \right] $
 \item Change in observation part of action:
 $\Delta A_{o,n} = \frac{R_o}{2} \left[ (\de_n')^2 - \de_n^2\right]$
 \item Total change in action: $\Delta A_n = \Delta A_{d,n} + \Delta A_{o,n}$
 \item Acceptance probability: $P_{acc} = \min(1, e^{-\Delta A_n})$
 \item If $U(0,1) < P_{acc}$ then accept the change: $\x_n' \to \x_n$ and $\F_n' \to \F_n$
\end{enumerate}
\item For each even $n$, a thread executes the same steps shown above
\item Perturb parameter $p_l$.  Launch $M+1$ threads to compute the resulting change in action, and then either accept or reject the change with the usual Metropolis rule.  Repeat for each parameter, $l=1,2,\ldots,N_p$ in sequence. 
\end{enumerate}
\caption{Pseudo-code for parallel state and parameter update.  For simplicity the matrices $\Ro$ and $\Rd$ are set be be $R_o$ and $R_d$ times the identity matrix, respectively. $U(a,b)$ is a random number between $a$ and $b$ drawn from a uniform distribution, and $\mathbf{U}$ is a vector of $D$ such components.}\label{stateUpdate}
\end{figure}

\begin{figure}
\noindent\makebox[\columnwidth]{ \includegraphics[width=0.6\columnwidth]{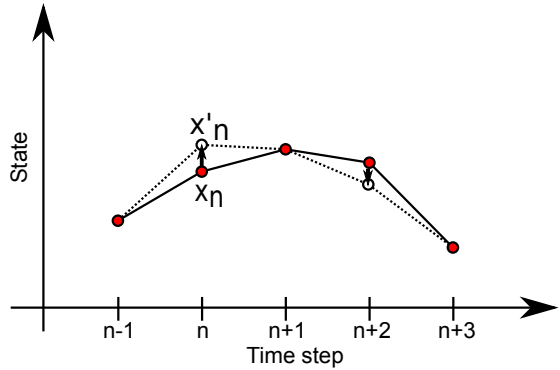}} 
\caption{Schematic of the state update process. All even time steps are updated simultaneously by different threads.  The solid circles show the current states (for illustration they are shown as scalars, but generally are vectors). The open circles show the proposed new states, and the dotted lines represent the dynamical terms in the action which would change.  After all even time steps have a chance to update, the same procedure is done to all the odd time steps.}
\label{diagram}
\end{figure}

\subsection{The general procedure}
First the current path $\X$ is set to an initial guess, and the observation time series $\Y$ is loaded from a file.  The path $\X$ includes the state vector $\x_n$ at every time step $n=0,1,\ldots, M$ , and the parameters.  The states and parameters are treated differently, because the parameters are forced to be time-independent but the states may vary in time.  The current path, the observation time series, the external drive signal (if present), and a running sum of moments of the path components are allocated in GPU memory and initialized with the appropriate data. 

Then the path update loop begins.  First the even $n$ states are updated, and then the odd $n$ states (see. Fig.~\ref{stateUpdate} and Fig.~\ref{diagram}).  The reason for doing the state update in two steps is to uncouple the state vectors:  to calculate the change in action due to perturbing $\x_n$, we need to know $\x_{n-1}$ and $\x_{n+1}$, but none of the other state vectors.  This way each even $n$, and then each odd $n$ can be updated independently, in any order or simultaneously.  

Once all the states have had a chance to be updated, then each parameter $p_l$ is given a chance to change in sequence, $l=1,2,\ldots, N_p$.  One parameter, $p_l$, is perturbed, and then $M+1$ threads are launched to calculate the change in action.  Each thread is assigned to one $n$, and it calculates the new RHS of the model differential equation $\F_n' = \F(\x_n')$.  From this information the total change in action is calculated, and one decision about whether to accept the proposed change or not is made using the usual Metropolis rule.  If the change is accepted, then $\F_n' \to \F_n$ for all $n$ and $p_l' \to p_l$.   

After all the states and parameters have had an opportunity to update, the current path can be used to update the path statistics.  Typically this will be the mean and variance of each component of the path, but covariances or higher moments could be of interest too. Another possibility is to record bin counts of some components of interest to make a histogram.  This step is skipped for the first $N_{init}$ path updates, and after that only done every $N_{skip}$-th path update.  The statistics collection happens on the GPU, also in parallel, and so the individual paths are not recorded.  This avoids costly data transfers between the GPU and CPU.       

\subsection{More details}

There are several more details which will now be discussed.  The first $N_{init}$ iterations are the initialization phase, during which no path statistics are recorded.  This phase is necessary to remove the influence of the initial guess path.  Two other things happen during this phase: simulated annealing, and automatic adjustment of the MC step sizes $\Delta_i$. 

The simulated annealing is done by putting a multiplier $\beta <1$ in front of the dynamics term of the action, $A_d$, and gradually increasing $\beta$ during the initialization phase, up to its final value of $\beta = 1$.  This way the action will initially be dominated by the quadratic observation term, $A_o$, which is smooth and has a single well-defined minimum.  Specifically, $\beta$ is initially set to $\beta = \beta_0 < 1$ and multiplied by the constant factor $f_\beta$ after each iteration, until $\beta =1$.  The constant multiplying factor is
\begin{equation}\label{fbeta}
f_\beta = (1/\beta_0)^{(1/N_{cool})}. 
\end{equation}

The $\beta$ plays a role similar to inverse temperature $T$ in a system with a Boltzmann distribution $P(\x) \propto \exp(-E(\x)/kT)$.  In our case the conditional path distribution is $P(\X|\Y) \propto \exp(-A_o(\X|\Y)) \exp(-\beta A_d(\X))$.  The two situations are not quite the same because in the second case there is a path-dependent and observation-dependent term, $A_o(\X|\Y)$ which is not multiplied by $\beta$.

An automatic procedure for adjusting the the MC step sizes, $\Delta_i$, is useful because there are typically many different step sizes which would be hard to tune by hand.  It is important to tune the sizes because if they are too small, then the proposed path changes will be very small.  They will be likely to be accepted, but very many iterations will be required for the path to change significantly. On the other hand,  if the step size is too large, most proposed changes will be rejected, and in this case also, many iterations will be required for the path to change significantly.  In this case there is a different step size for each component of the path indexed by $i$.  The delta adjustment rule used was 
\[
\Delta_i \leftarrow \Delta_i \left[ 1 + \alpha \left(\frac{N_{acc,i}}{N_{skip}} - f_{acc} \right) \right],
\] 
where $N_{acc,i}$ is the number of accepted changes for component $i$ over the past $N_{skip}$ number of iterations, $f_{acc}$ is the target acceptance rate, and $\alpha$ is a constant that controls the adjustment rate.  For a different approach which utilizes a proportional integral controller to tune step sizes see \cite{banfelder}.  This rule is applied every $N_{skip}$-th iteration, and only during the initialization phase.  It is important to note that adjusting the step sizes will bias the distribution, so samples generated during the adjustment phase should not be used when calculating statistics.  It is run in parallel on the GPU, with one thread assigned to each path component.   It was shown in \cite{roberts01, gelman96} that the optimal value for the acceptance rate is 0.23 in the limit of a infinite dimensional multivariate normal target distribution, so that is the target acceptance rate that we use here (see also \cite{kincaid}). It is important that each parameter and state variable are tuned independently, because the scales may be very different. 

Another useful thing to do is to set boundaries on parameters.  This can be done by adding a large penalty to the action when a proposed change would move the parameter outside of the boundary.  This makes the proposed change be rejected.  This is one way of incorporating a prior distribution of parameters, in the simple case where the distribution is uniform with the bounds and zero outside of the bounds.  

The random number generation is done in parallel using the ``Hybrid Tausworthe'' algorithm described in \cite{GPUGems3}. Each thread is given its own initial seed,  and so each thread will generate a different pseudo-random sequence. To test the quality of the random number generator we generated 512 different sequences in parallel of 10,000 random numbers each, and then computed the cross-correlations and autocorrelations.  All correlations came out between -0.04 and 0.04, except the autocorrelations at zero time lag, which was one by choice of normalization. 
\subsection{GPU-specific details}

There are several important limitations to keep in mind when designing CUDA code.  The threads are grouped into blocks which run together.  The amount of threads allowed in a block depends on how many resources each thread uses (often limited by shared memory usage or by number of registers used), but its maximum possible value is 1024 threads per block.  The threads within a block can communicate with each other through shared memory, and can synchronize with each other.  In contrast, threads in different blocks can have no interaction, and may be run in any order or simultaneously, so the code must be designed such that the order of block execution does not matter.  

The GPUs have a large amount of global memory which is accessible to all threads, but has relatively high latency.  The CPU can read or write to the global GPU memory, but this process should be kept to a minimum, because this process is relatively slow.  Each block has a small amount of shared memory (49KB on the NVIDIA GTX 460 for example) which is much faster. The shared memory should be used where possible for temporary storage within the kernel call, but since its size is limited, there is often a trade off between using shared memory and being able to run more threads per block. 

A common approach, which was used in the state update kernel, is to load the necessary pieces of information from global memory into shared memory at the beginning of the kernel.  In this case the data was the $\F_n$ values and the $\x_n$ values for the relevant time steps.  There is a overlap of one time step on the data which is read (but not of the data which is changed) between each block to make the neighboring values accessible.  Since the blocks cannot synchronize with each other, it is necessary to do the even and odd $n$ values in two separate kernel invocations.  This ensures that all the even $n$ finish before any odd $n$ start. 

The update process for the parameters is slightly more tricky, because in this case only one decision about whether to accept or reject is made, but it is based on the calculation of several blocks.  This is handled by having each block calculate a change in action due to its assigned time steps.  Each thread within the block calculates a change in action, and then participates in a parallel sum to get a total change in action for the whole block.  Then after all the blocks finish another kernel is called which runs on a single thread and sums the block sums and decides whether to accept the change.  If the change is accepted, the parameter is updated, and then the $\F_n$ values are updated.  Actually two copies of the $\F_n$'s are kept in global memory, one which was calculated from the current path and one which was calculated from the proposed path.  There is a pointer to select which copy is current, and this pointer is flipped when the change is accepted, so that large memory transfers are avoided.   

The observation time series and external drive signal (if present) are loaded into constant memory, which is a small (65KB on NVIDIA GTX 460) memory which can be quickly read by the GPU but not written to. For longer time series, this memory may be too small, in which case global memory is used instead.  Also the data is organized in memory so that threads with contiguous indices will access contiguous regions of memory.  This allows the memory transfers to be coalesced and happen more efficiently.  All calculations were done using single-precision floating point arithmetic.    

\section{Results}

\subsection{Example problem: Hodgkin-Huxley}
We now discuss the Hodgkin-Huxley neuronal model as an example problem.   For other approaches to the similar problems see \cite{huys,rod}.  The model, first developed in 1952 based on experiments on the squid giant axon \cite{HH}, treats the cell membrane as a capacitative layer with three types of conducting channels.  The conductance of the sodium and potassium ion channels depends on the voltage across the membrane $V(t)$.  The differential equation for voltage is:

\begin{align}\label{dVdt}
\frac{dV}{dt} &= p_1 I_{stim}(t) \notag \\
&+ p_2 m^3(t) h(t) \left( p_3- V(t) \right) \notag \\
&+ p_4 n^4(t) \left( p_5 -V(t) \right) \notag \\
&+ p_6 \left( p_7 -V(t) \right).
\end{align}

The first term is an external stimulus current injected into the cell, the second term is the sodium ion current, the third term is potassium ion current, and the last term is a leak current.  The voltage dependence of the conducting channels is modeled through gating probabilities $m(t), n(t), h(t)$, all between zero and one.  The dynamics of the gating probabilities is specified by  first-order kinetic equations with opening and closing rates which are functions of $V$.  Equivalently, the equations can be expressed in terms of a steady state value $a_\infty(V)$ and a time constant $\tau_a(V)$, where $a$ stands for $m,n,$ or $h$:
 
\begin{align}\label{dadt}
\frac{da}{dt} &= \frac{a_{\infty}(V) - a(t)}{\tau_a(V)}.
\end{align}

Phenomenological analytic expressions for these functions were discussed empirically by Hodgkin and Huxley.  For this example we use different but qualitatively similar functions:
\begin{align}
a_{\infty}(V) &= \frac{1}{2} + \frac{1}{2} \tanh\left(\frac{V-V_a}{\Delta V_a} \right) \\
\tau_a(V) &= \tau_{a0} + \tau_{a1} \left(1 - \tanh^2\left(\frac{V-V_a}{\Delta V_a} \right) \right).
\end{align}

\begin{figure}
\noindent\makebox[\columnwidth]{ \includegraphics[width=0.7\columnwidth]{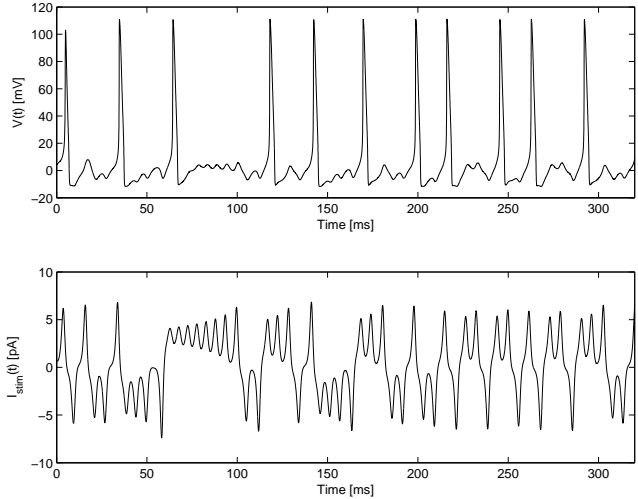}} 
\caption{The simulated voltage measurement and the chaotic stimulus current.  These two time series plus the Hodgkin- Huxley model (with seven unknown parameters) are inputs into the PIMC procedure.  The $V(t)$ data is 8000 discrete points (although it is shown as a solid line) sampled at 25 kHz.}
\label{inputs}
\end{figure}

Simulated data was created by integrating the Hodgkin-Huxley model, with a chaotic stimulus current generated from one of the variables of the Lorenz model \cite{lorenz63}.  Values for the parameters in the gating equation were chosen to qualitatively match what was found by Hodgkin and Huxley, and are given in Table~\ref{kineticParams}. Simulated measurements of transmembrane voltage $V(t)$ were generated from the model output with a sampling frequency of 25 kHz.  This simulated data along with the chaotic stimulus current is shown in Fig.~\ref{inputs}. 

The model has a total of four state variables, $V, n,m$, and $h$, but only the voltage is observed.  This corresponds to the typical situation in real experiments.  In a current clamp experiment on a single isolated neuron, a known current is injected into the cell while the transmembrane voltage response is recorded.  In this version of the model,  seven numbers $p_1,p_2, \ldots, p_7$, were treated as unknown parameters, and all other constants were fixed at the same values for data generation and for the analysis.  One can determine all of the constants in $a_{\infty}(V)$ and $\tau_sa(V)$ with the same methods. Our focus is on the computational procedures here. In the present case the observation function $\h(\x_n)$ simply returns the the voltage component of $\x_n$, but in general it could be any function of $\x_n$.   

In the example calculation, one million path update iterations were performed, with $N_{skip} = 400$.  The first half million were excluded from the statistics ($N_{init} = 500,000$). Simulated annealing was done over the first $N_{cool} = 100,000$ iterations, with $\beta_0 = 0.01$, and $\beta$ was incremented by multiplying by $f_\beta$  (Eq.~\ref{fbeta}) after each iteration.  The target acceptance rate was $f_{acc}=0.23$ and the step size adjustment rate was $\alpha = 0.02$.  The initial settings for Monte Carlo step sizes were $\Delta_i = 2 \times 10^{-3}$ for voltage and $\Delta_i = 10^{-3}$ for the $n,m,$ and $h$ state variables. 

\begin{figure}
\noindent\makebox[\columnwidth]{ \includegraphics[width=0.7\columnwidth]{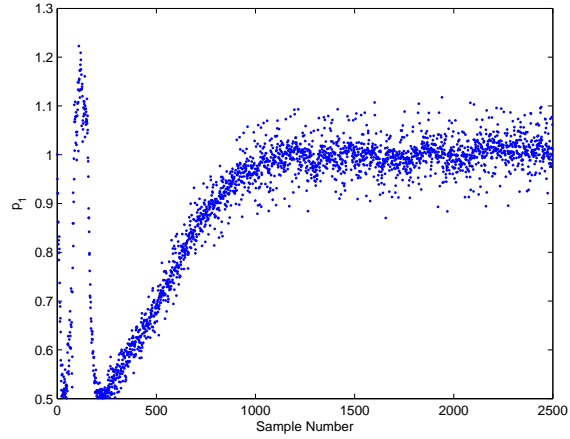}} 
\caption{The evolution of one of the parameters as the MC path update procedure runs.  The first 1250 samples where discarded.  Simulated annealing was applied during the first 250 iterations.  Eventually the parameter settles down and fluctuates around a mean of 1.02.}
\label{pEvo}
\end{figure}

\begin{figure}
\noindent\makebox[\columnwidth]{ \includegraphics[width=0.7\columnwidth]{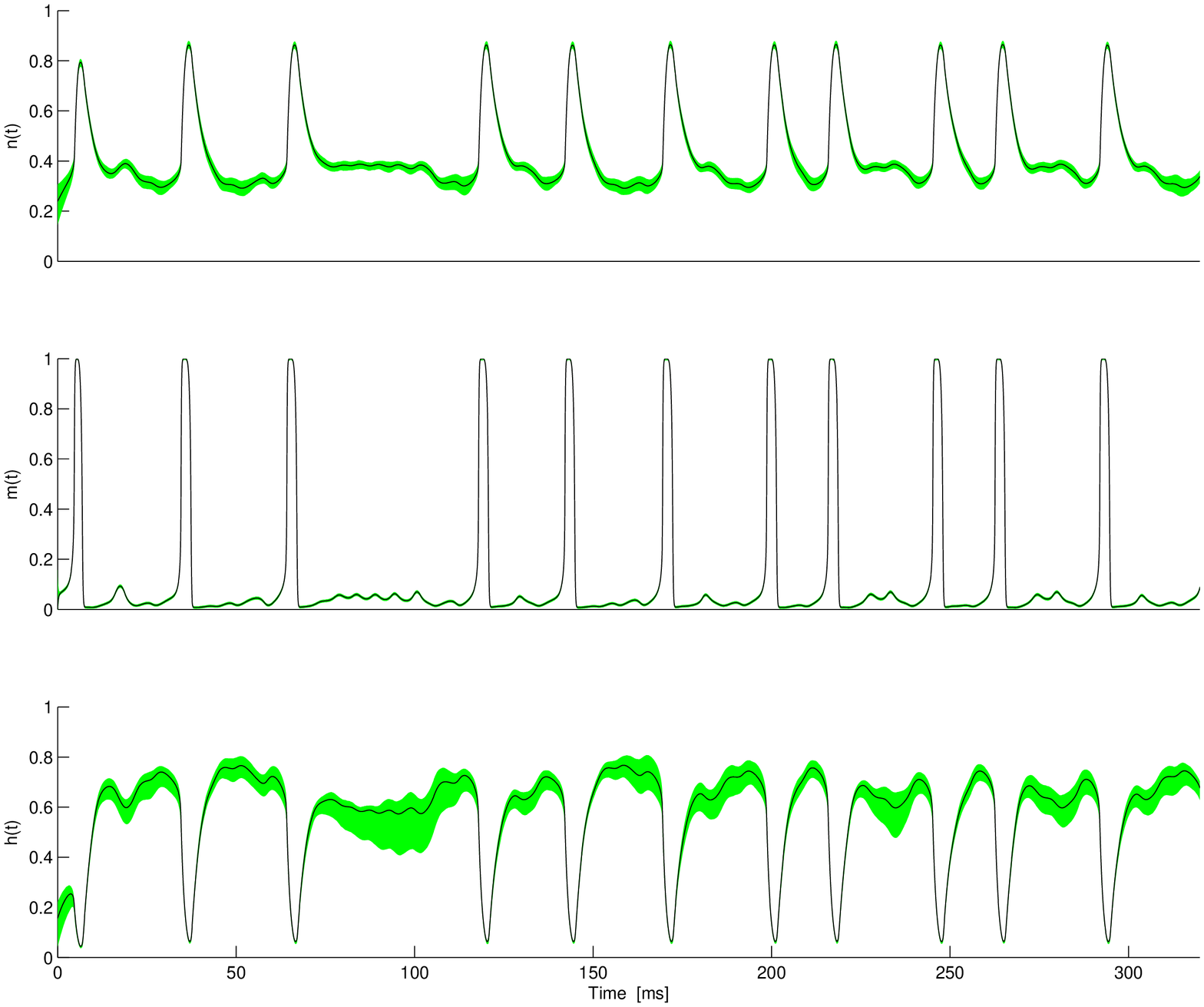}} 
\caption{The output of the PIMC process.  These are the three unobserved states.  The black line is the true value (which is known because the data was simulated). The solid region is centered on the mean and extends to plus or minus 4 standard deviations from the mean. }
\label{unobsStates}
\end{figure}

The number of data points used was $M=8000$, or 320 ms of data.  The matrix $\Rd$ was set to be $[\Rd]_{V,V}= 100$, $[\Rd]_{a,a} = 10^6$ ($a \in \{n,m,h\}$),  and all other components zero.  Since the observation is one-dimensional $\Ro$ is a scalar, and was set to 100.  Figure \ref{pEvo} shows the evolution of one parameter, $p_1$, to give an idea of the equilibration process.  The mean and standard deviation of each component of the path, that is each state at each time step and each parameter, was calculated.  The mean and the standard deviation of the unobserved states shown in Fig.~\ref{unobsStates}.  The estimated parameters values are given in Table ~\ref{paramsTable}.  This calculation took 683 seconds to run for one million iterations on an NVIDIA GTX 460 GPU which has 224 cores. 

\begin{table}
\begin{center}\caption{Constants used in the kinetic equations.}
\begin{tabular} {|l|c|c|c|c|}
\hline
Name			&	 $V_a$ [mV]	&	$\Delta V_a$ [mV] 	&	$\tau_{a0}$[ms]&	$\tau_{a1}$ [ms]\\
\hline
$n$, K Activation &	10			&	30				&	1.0			& 5.0		\\
$m$, Na Activation &	25			&	15				&	0.1			& 0.4	\\
$h$, Na Inactivation & 5			&	-15				&	1.0			& 7.0\\
\hline
\end{tabular}
\label{kineticParams}
\end{center}
\end{table}

\begin{table}
\begin{center}\caption{Estimated parameters.}
\begin{tabular} {|c|c|c|c|c|}
\hline
Name			&	Mean 		& St. Dev.  & Actual  & Units\\
\hline
$p_1$			& 1.02			& 0.04		& 1  & 1/pF\\
$p_2	$		& 127			& 2			& 120 & 1/ms\\
$p_3$			& 115			& 0.1		& 115 & mV \\
$p_4$			& 20.3			& 0.3		& 20 & 1/ms \\
$p_5$			& -12.0			& 0.08		& -12 & mV \\
$p_6$			& 0.33			& 0.03		& 0.3 & 1/ms\\
$p_7$			& 9.3			& 1.0		& 10.6 & mV\\
\hline
\end{tabular}
\label{paramsTable}
\end{center}
\end{table}

\begin{figure}
\noindent\makebox[\columnwidth]{ \includegraphics[width=0.6\columnwidth]{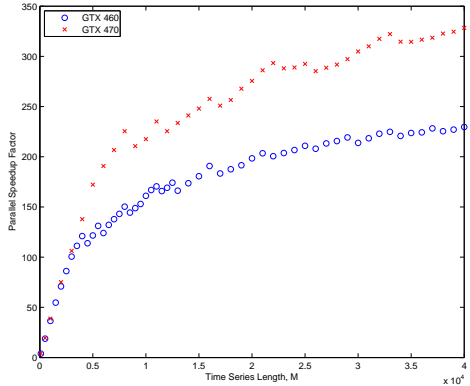}} 
\caption{The parallel speedup factor as a function of time series length M, using two different NVIDIA GPUs.  The parallel speedup factor is defined as $T_{CPU}/T_{GPU}$.  The performance of the two GPUs is about the same for small $M$, where they are underutilized, but as $M$ increases, the GTX 470 performs better because it has more cores.}
\label{speedup}
\end{figure}

For timing purposes, the same calculation was done over a range of time series lengths from $M=100$ up to $M=40,000$, but with 1/10th as many iterations, and the execution time, $T_{GPU}$, was recorded.   The same calculation was done on two different NVIDIA GPUs, the GTX 460 with 224 cores and the GTX 470 with 448 cores. In all cases, the number of threads per block was set to 100.  For comparison a similar, but not exactly the same, calculation was done on a single core of an Intel Core i3 CPU.  A linear time scaling of $T_{CPU} =M \times (1.24$ s) was fit from several trials on the CPU.  In Fig.~\ref{speedup} the parallel speedup factor, $T_{CPU}/T_{GPU}$, is displayed.   

\section{Discussion}
Running the PIMC in parallel decreased the computation time by a factor of up to 200 or 300, depending on which GPU was used.  As the time series becomes longer, more of the GPU cores are simultaneously utilized.  The specifics of how the speedup factor varies as a function of time series length depends on the particular GPU, and on how CUDA schedules the threads to utilize the available resources.      

One way to assess the quality of the model is to use the model to integrate the final state at time $t=T$ forward in time and compare to additional observations.  One possibility is to use the mean final state $\left<x_T \right>$ and mean value of the parameters, $\left<p_l \right>$, to generate a single trajectory, which can then be compared to the observed time series.  However, this may not be adequate when the dynamics is noisy, or when the model is chaotic.  Another approach is to generate an ensemble forecast, with the appropriate amount of noise added into the model integration.  Many noisy model integrations should be done, by using many different final states and parameter values as generated by the Monte Carlo procedure, and many different noise sequences.  Then the ensemble forecast can be compared to the single observed trajectory in a statistical way.  

Another way to asses the quality of the model is to look at the model errors, $\ep_n$, and the observation errors, $\de_n$.  These will have specific values for each Monte Carlo path generated, since they are functions of the path, $\X$.  If the model is correct, then the statistics of these quantities should be consistent with our assumptions about the noise.  In the example problem the assumptions are that both noises are Gaussian and  white.  Therefore the means should be zero, the variances should come out as specified by the $\Ro$ and $\Rd$ matrices, and there should be no correlation in time or among the different components.  In the example problem presented here, there is a source of model error which comes from the discretization of the model differential equations.  This appears as deviations of $\left<\ep_n \right>$ from zero at the times when $V(t)$ spikes. 

The method presented in this paper is quite general, and can be applied to any dynamical system.  All that is needed is a time series of observations, a Markov model of the dynamics, and some assumptions about the noise in the dynamics and the noise in the observations.  By utilizing GPU technology, it becomes practical to run the method on realistic problems using desktop computers.    

\section*{\bf Acknowledgments} We thank Marius Buibas and Reza Farsian for useful discussions about parallelizing the algorithm.  Support from the US Department of Energy (Grant DE-SC0002349 ) and the National Science Foundation (Grants  IOS-0905076 and PHY-0961153) are gratefully acknowledged. Partial support from the NSF sponsored Center for Theoretical Biological Physics is also appreciated. 


\input{quinngpu11-D.bbl}
\end{document}

%% file: definitions.tex
\def\Y{\mathbf{Y}}
\def\X{\mathbf{X}}

\def\F{\mathbf{F}}

\def\y{\mathbf{y}}
\def\x{\mathbf{x}}
\def\p{\mathbf{p}}
\def\h{\mathbf{h}}

\def\ep{\bm{\epsilon}}
\def\de{\bm{\delta}}

\def\Ro{\mathbf{R_o}}
\def\Rd{\mathbf{R_d}}
\def\N{\mathcal{N}}

\newcommand{\be}{\begin{equation}}
\newcommand{\ee}{\end{equation}}
\newcommand{\bea}{\begin{eqnarray}}
\newcommand{\eea}{\end{eqnarray}}

%% file: quinngpu11-D.bbl
\begin{thebibliography}{10}
\expandafter\ifx\csname url\endcsname\relax
  \def\url#1{\texttt{#1}}\fi
\expandafter\ifx\csname urlprefix\endcsname\relax\def\urlprefix{URL }\fi
\expandafter\ifx\csname href\endcsname\relax
  \def\href#1#2{#2} \def\path#1{#1}\fi

\bibitem{cox64}
H.~Cox, On the estimation of state variables and parameters for noisy dynamic
  systems, IEEE transactions on automatic control (1964) 5--12.

\bibitem{friedland66}
B.~Friedland, I.~Bernstein, Estimation of the state of a non-linear process in
  the presence of nongaussian noise and disturbances, Journal of the Franklin
  Institute 281~(6).

\bibitem{kalman60}
R.~Kalman, A new approach to linear filter and prediction problems, J. Basic
  Eng. 82 (1960) 35--45.

\bibitem{vanLeeuwen09}
P.~J.~V. Leeuwen, Particle filtering in geophysical systems, Monthly Weather
  Review 137 (2009) 4089--4114.

\bibitem{arulampalam}
S.~Arulampalam, S.~Maskell, N.~Gordon, T.~Clapp, A tutorial on particle filters
  for on-line non-linear/non-{G}aussian {B}ayesian tracking, IEEE Transactions
  on Signal Processing 50 (2001) 174--188.

\bibitem{doucet}
A.~Doucet, S.~Godsill, C.~Andrie, On sequential {M}onte {C}arlo sampling
  methods for {B}ayesian filtering, Statistics and Computing 10 (2000)
  197--208.

\bibitem{eyink05}
F.~J. Alexander, G.~L. Eyink, J.~M. Restrepo, Accelerated {M}onte {C}arlo for
  optimal estimation of time series, Journal of Statistical Physics 119~(5).

\bibitem{restrepo}
J.~M. Restrepo, A path integral method for data assimilation, Physica D:
  Nonlinear Phenomena 237~(1) (2008) 14 -- 27.
\newblock \href {http://dx.doi.org/DOI: 10.1016/j.physd.2007.07.020}
  {\path{doi:DOI: 10.1016/j.physd.2007.07.020}}.

\bibitem{apte07}
A.~Apte, M.~Hairer, A.~Stuart, J.~Voss, Sampling the posterior: An approach to
  non-{G}aussian data assimilation, Physica D: Nonlinear Phenomena 230~(1-2)
  (2007) 50 -- 64.
\newblock \href {http://dx.doi.org/DOI: 10.1016/j.physd.2006.06.009}
  {\path{doi:DOI: 10.1016/j.physd.2006.06.009}}.

\bibitem{abar09}
H.~D. Abarbanel, Effective actions for statistical data assimilation, Physics
  Letters A 373~(44) (2009) 4044 -- 4048.
\newblock \href {http://dx.doi.org/DOI: 10.1016/j.physleta.2009.08.072}
  {\path{doi:DOI: 10.1016/j.physleta.2009.08.072}}.

\bibitem{preis09}
T.~Preis, P.~Virnau, W.~Paul, J.~J. Schneider, {GPU} accelerated {M}onte
  {C}arlo simulation of the 2d and 3d {I}sing model, Journal of Computational
  Physics 228~(12) (2009) 4468 -- 4477.

\bibitem{pham}
D.~T. Pham, Stochastic methods for sequential data assimilation in strongly
  nonlinear systems, Mon. Weath. Rev. 129 (2001) 1194--1207.

\bibitem{jcq10}
J.~C. Quinn, H.~D. Abarbanel, State and parameter estimation using {M}onte
  {C}arlo evaluation of path integrals, Quarterly Journal of the Royal
  Meterological Society 136~(652) (2010) 1855--1867.

\bibitem{mackay}
D.~J. MacKay, Information Theory, Inference, and Learning Algorithms, Cambridge
  University Press, 2003.

\bibitem{zakai}
M.~Zakai, On the optimal filtering of diffusion processes, Probability Theory
  and Related Fields 11 (1969) 230--243.

\bibitem{MRT}
N.~Metropolis, A.~Rosenbluth, M.~Rosenbluth, A.~Teller, E.~Teller, Equations of
  state calculations by fast computing machines, J. Chem. Phys 21~(1953).

\bibitem{neal}
R.~M. Neal, Probabilistic inference using {M}arkov chain {M}onte {C}arlo
  methods, Technical Report CRG-TR-93-1.

\bibitem{hastings}
W.~K. Hastings, Monte {C}arlo sampling methods using {M}arkov chains and their
  applications, Biometrika 57~(1) (1970) 97--109.

\bibitem{pangali}
C.~Pangali, M.~Rao, B.~Berne, On a novel monte carlo scheme for simulating
  water and aqueous solutions, Chemical Physics Letters 55~(3) (1978) 413 --
  417.
\newblock \href {http://dx.doi.org/DOI: 10.1016/0009-2614(78)84003-2}
  {\path{doi:DOI: 10.1016/0009-2614(78)84003-2}}.

\bibitem{banfelder}
J.~R. Banfelder, J.~A. Speidel, M.~Mezei, Automatic determination of stepsize
  parameters in monte carlo simulation tested on a bromodomain-binding
  octapeptide, Algorithms 2~(1) (2009) 215--226.
\newblock \href {http://dx.doi.org/10.3390/a2010215}
  {\path{doi:10.3390/a2010215}}.

\bibitem{roberts01}
G.~Roberts, J.~Rosenthal, Optimal scaling for various {M}etropolis-{H}astings
  algorithms, Statistical Science 16~(4) (2001) 351--367.

\bibitem{gelman96}
A.~Gelman, G.~Roberts, W.~Gilks, Efficient {M}etropolis jumping rules, Bayesian
  Statistics (1996) 599--607.

\bibitem{kincaid}
R.~H. Kincaid, H.~A. Scheraga, Acceleration of convergence in monte carlo
  simulations of aqueous solutions using the metropolis algorithm. hydrophobic
  hydration of methane, Journal of Computational Chemistry 3~(4) (1982)
  525--547.
\newblock \href {http://dx.doi.org/10.1002/jcc.540030410}
  {\path{doi:10.1002/jcc.540030410}}.

\bibitem{GPUGems3}
H.~Nguyen, GPU Gems 3, NVIDIA Corporation, 2008.

\bibitem{huys}
Q.~J.~M. Huys, M.~B. Ahrens, L.~Paninski, {Efficient Estimation of Detailed
  Single-Neuron Models}, J Neurophysiol 96~(2) (2006) 872--890.
\newblock \href {http://dx.doi.org/10.1152/jn.00079.2006}
  {\path{doi:10.1152/jn.00079.2006}}.

\bibitem{rod}
M.~Rodriguez-Fernandez, J.~A. Egea, J.~R. Banga, Novel metaheuristic for
  parameter estimation in nonlinear dynamic biological systems, BMC
  Bioinformatics 7 (2006) 483--501.
\newblock \href {http://dx.doi.org/10.1186/1471-2105-7-483}
  {\path{doi:10.1186/1471-2105-7-483}}.

\bibitem{HH}
A.~Hodgkin, A.~Huxley, A quantitative description of membrane current and its
  application to conduction and excitation in nerve, J. Physiology (1952)
  500--544.

\bibitem{lorenz63}
E.~N. Lorenz, Deterministic nonperiodic flow, Journal of the Atmospheric
  Sciences 20 (1963) 130--141.

\end{thebibliography}
